\input{aipcheck.tex}

\renewcommand{\vec}[1]{\mathbf{#1}}

\newcommand\sps{\space\space\space\space}
\documentclass[final]{aipproc}

\layoutstyle{6x9}

\SetInternalRegister\hbadness{8000} 

\newcommand\doingARLO[2][]{%
  \ifx\mmref\undefined #1\else #2\fi
}

\begin{document}
\title[Residual energy in MHD turbulence]{Residual energy in MHD turbulence and in the solar wind}


\author{Stanislav Boldyrev}{
  address={Department of Physics, University of Wisconsin, Madison, WI 53706, USA}, 
}
\author{Jean Carlos Perez}{
  address={Department of Physics, University of Wisconsin, Madison, WI 53706, USA},
 altaddress={Space Science Center, University of New Hampshire, Durham, NH 03824, USA} 
}
\author{Vladimir Zhdankin}{
  address={Department of Physics, University of Wisconsin, Madison, WI 53706, USA},
}



\copyrightyear{2011}

\begin{abstract}
Recent observations indicate that kinetic and magnetic energies are not in equipartition in the solar wind turbulence. Rather, magnetic fluctuations are more energetic and have somewhat steeper energy spectrum compared to the velocity fluctuations. This leads to the presence of the so-called residual energy $E_r=E_v-E_b$ in the inertial interval of turbulence. This puzzling effect is addressed in the present paper in the framework of weak turbulence theory. Using a simple model of weakly colliding Alfv\'en waves, we demonstrate that the kinetic-magnetic equipartition indeed gets broken as a result of nonlinear interaction of Alfv\'en waves. We establish that magnetic energy is indeed generated more efficiently as a result of these interactions, which proposes an explanation for the solar wind observations.   \end{abstract}

\date{\today}

\maketitle

\section{Introduction}


Magnetohydrodynamics (MHD) provides possibly the simplest framework for describing turbulence observed in the solar wind. When compressibility and the dissipation effects may be neglected, MHD equations have exact solutions in the form of Alfv\'en waves that propagate (in both directions) along the background magnetic field with the same constant Alfv\'en speed. The characteristic feature of those Alfv\'en wave packets is that their magnetic and velocity perturbations, ${\bf b}({\bf x}, t)$ and ${\bf v}({\bf x}, t)$, are either exactly aligned or counteraligned with each other:  for wave packets propagating against the direction of the background field, one has ${\bf v}({\bf x},t)={\bf b}({\bf x},t)$, for wave packets propagating in the direction of the field, ${\bf v}({\bf x},t)=-{\bf b}({\bf x},t)$. Counterpropagating Alfv\'en wave packets interact when they overlap, or collide, with each other, e.g. \cite{goldreich_s95,biskamp03}. 

The symmetry between magnetic and velocity perturbations in individual Alfv\'en waves may suggest that magnetic and kinetic energies should be in equipartition in the developed MHD turbulence. Indeed, phenomenological theories of MHD turbulence often do not distinguish between the two spectra, assuming (explicitly or implicitly) that they are the same.   However, recent observational data indicate that kinetic and magnetic fluctuations in the solar wind turbulence are not in exact equipartition at 1AU. There is slight excess of magnetic energy and  magnetic fluctuations have slightly steeper Fourier energy spectrum, e.g., \cite{podesta_rg07,tessein_etal09,podesta_b09,chen_etal10,borovsky11,chen_etal11}. A possible explanation may be that the solar wind turbulence is in a transient state at 1AU \cite{roberts10}. Another possibility is that the solar wind contains some remnant structures, magnetic field discontinuities or current sheets, advected from the sun and not solely generated by the turbulence \cite{borovsky10,li_etal11}. Those explanations are interesting and may be plausible. 

\begin{figure}
  \includegraphics[height=.3\textheight]{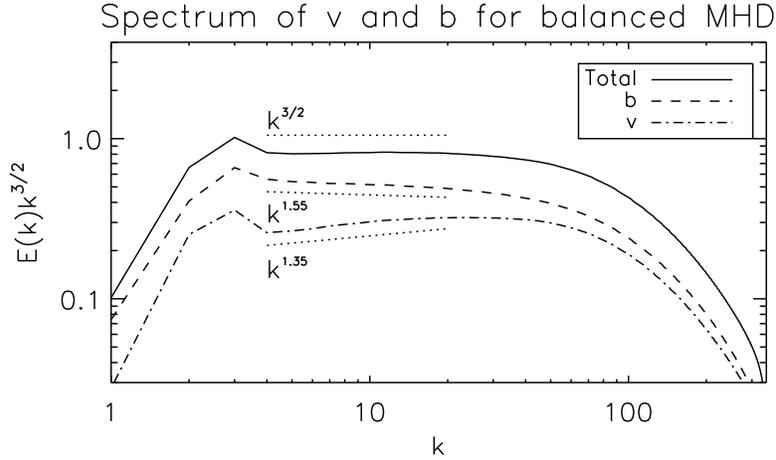}
  \caption{The field-perpendicular Fourier spectra of kinetic, magnetic, and total energies in numerical simulations of strong incompressible MHD turbulence with a strong guide field. The Elsasser fields $z^+$ and $z^-$ are driven independently at large scales, so that no residual energy is generated by the driving mechanism. The numerical resolution is $1024^2 \times 256$, the Reynolds number is $Re \approx 5600$. The details of the simulations can be found in \cite{boldyrev_pbp11}. }
  \label{fig}
\end{figure}

Quite interestingly, recently it has been found that incompressible steady-state MHD turbulence exhibits a very similar mismatch between the kinetic and magnetic energies to that observed in the solar wind \cite{boldyrev_pbp11}. Moreover, it has been discovered that mismatch between magnetic and kinetic energies exists even in the case of {\em weak} MHD turbulence, that is, turbulence of mostly independent Alfv\'en waves where equipartition between kinetic and magnetic energies is generally expected \cite{boldyrev_p09,wang11}. These developments suggest that breakdown of kinetic-magnetic equipartition may, in fact, be  a fundamental property of Alfv\'enic turbulence (rather than an artifact or some nonuniversal plasma effects), which can be understood in the framework of incompressible MHD (see Fig.~\ref{fig}). This effect may be complementary to the previously proposed explanations for the solar wind observations \cite{roberts10,borovsky10,li_etal11}.  

The difference between kinetic and magnetic energies can be characterized by the so-called residual energy, $E_r=\int (v^2-b^2)d^3x=\int ({\bf z}^+\cdot {\bf z}^-) d^3x$, where ${\bf z}^{\pm}={\bf v}\pm{\bf b}$. Early studies of MHD turbulence based on some closure assumptions indeed indicated that kinetic and magnetic energies can be different \cite{grappin_pl83,zank_ms96,muller_g05}. However, the question remains relatively poorly explored since most of the models of MHD turbulence are based on the quantities like total energy $E=\frac{1}{2}\int (v^2+b^2)d^3x$ and cross-helicity $H_c=\int ({\bf v}\cdot {\bf b})d^3x$, or their combinations, the so-called Elsasser energies $E^{\pm}=\int (z^{\pm})^2 d^3x$, neither of which contains the information about the residual energy. In what follows we present a simple model illustrating how residual energy is generated in a fundamental nonlinear process of MHD turbulence -- collision of counter-propagating Alfv\'en waves.

\section{A model for residual energy generation}
To formulate the model, we use the standard representation of the incompressible MHD equations in terms of the Els\"asser variables, 
\begin{equation}
  \left( \frac{\partial}{\partial t}\mp\vec v_A\cdot\nabla\right)\vec
  z^\pm+\left(\vec z^\mp\cdot\nabla\right)\vec z^\pm = -\nabla P,
  \label{mhd-elsasser}
\end{equation}
where the Els\"asser variables are defined as $\vec z^\pm=\vec
v\pm\vec b$. In these formulas, $\vec v$ is the fluctuating plasma velocity (the equations are written in a frame with zero mean flow velocity), $\vec b$ is
the fluctuating magnetic field normalized by $\sqrt{4 \pi \rho_0}$,
${\bf v}_A={\bf B}_0/\sqrt{4\pi \rho_0}$ is the Alfv\'en velocity corresponding to the
uniform magnetic field ${\bf B}_0$ (so that the total magnetic field is ${\bf B}={\bf B}_0+{\bf b}$),  $P=p/\rho_0+b^2/2$ is the total pressure that includes the
plasma pressure, $p$, and the magnetic pressure, $\rho_0$ is the constant 
mass density, and we neglect driving and dissipation terms, e.g. \cite{biskamp03,marsch_m87}. 

An important fact can be derived from Eq.~(\ref{mhd-elsasser}). If we consider field  configurations where $\vec z^\mp(\vec x,t)\equiv 0$, then an arbitrary
function $\vec z^\pm(\vec x,t)=F^\pm(\vec x\pm\vec v_At)$ is an exact
solution of the nonlinear equations~(\ref{mhd-elsasser}). This solution represents a non-dispersive Alfv\'en  wave propagating
along the direction of the guide field with the velocity $\mp\vec v_A$. Nonlinear interactions are thus the result
of interactions between counter-propagating Alfv\'en wave packets. If during one interaction (one crossing time) the deformation of the wave packets is significant, then the turbulence is called strong; if the deformation is weak, the turbulence is called weak. Weak turbulence can 
be studied using perturbation theory, e.g., \cite{ng_b96,galtier_nnp00}. In what follows we study how weak interactions of counter-propagating Alfv\'en wave packets can generate residual energy, even if the residual energy is absent originally.  
 
Let us assume that the strong guide magnetic field ${\bf B}_0$ is in $z$-direction. We consider our system in a box that is periodic in x and y directions, with the dimensions $2\pi\times 2\pi$, and sufficiently long in the $z$-direction.  When the interaction is neglected, we assume that the waves are shear-Alfv\'en waves with polarizations normal to the background field, and that they have the following simple field-perpendicular structures:
\begin{eqnarray}
{\bf z}^+_0(x,y,z,t)=f^+(z+v_At)\cos(y){\hat{\bf x}}, \label{zero+} \\
{\bf z}^-_0(x,y,z,t)=f^-(z-v_At)\cos(x){\hat{\bf y}},
\label{zero-}
\end{eqnarray}
The functions (\ref{zero+}, \ref{zero-}), supplemented by $P_0\equiv \mbox{const}$ that follows from the incompressibility condition, are the exact solutions of Eq.~(\ref{mhd-elsasser}) without the nonlinear terms. This form ensures that without the nonlinear interaction, the residual energy is identically zero: ${\bf z}_0^+\cdot {\bf z}_0^-=0$. For simplicity, we also assume that the waves propagating in the opposite directions are not correlated with each other, that is, $\langle f^+f^-\rangle =0$. This happens, for example, if the functions $f^+$ and $f^-$ are random functions of their arguments with certain correlation  lengths $l_{\|}$, which can be interpreted as the typical size of the interacting wave packets in the direction of the guide magnetic field ${\bf B}_0$.  The angular brackets then mean averaging over the  $z$-direction (practically, over scales much larger than $l_\|$).

We will consider the case of weak turbulence, that is, $z_0^\pm l_\| \ll v_A l_\perp$, where $l_\perp=2\pi$ is the field-perpendicular size of the interacting wave packets. This condition follows from the fact that the linear terms in Eq.~(\ref{mhd-elsasser}), estimated as $v_A z_0^\pm /l_\|$, must be larger than the nonlinear ones, $z_0^\pm z_0^\mp/l_\perp$. In addition we assume that $l_\| \gg l_\perp$, which allows us to retain only the field-perpendicular gradient of the pressure term, and ensures that only shear-Alfv\'en waves are generated by  nonlinear interactions, see, e.g., \cite{galtier_nnp00,galtier_c06,perez_b09}. Let us now assume that interaction is switched on at time $t=0$. The solution can then be found  perturbatively in small parameter $\epsilon=z^\pm/v_A$, that is, we can write: 
\begin{eqnarray}
{\bf z}^\pm={\bf z}^\pm_0+{\bf z}^\pm_1+{\bf z}^\pm_2+\dots \label{expansion}
\end{eqnarray} 
We illustrate the main steps of the derivation using the field ${\bf z}^+$. It is convenient to introduce the new variables $\xi=z+v_At$ and $\tau=t$; then the equation for ${\bf z}_1^+$ takes the form:
\begin{eqnarray}
\partial_\tau {\bf z}^+_1=-\left(\vec z_0^-\cdot\nabla\right)\vec z_0^+ -\nabla_\perp P_1.
\label{first+}
\end{eqnarray}
The pressure $P_1$ should ensure incompressibility of ${\bf z}^+_1$; it can be found by taking the divergence of both sides of (\ref{first+}). One finds this way that
\begin{eqnarray}
P_1=\frac{1}{2}f^+f^-\sin(x)\sin(y).
\end{eqnarray}
We can now integrate Eq.~(\ref{first+}) with respect to time, keeping $\xi=const$, and get:
\begin{eqnarray}
{\bf z}^+_1=G^+(z,t)\left[ \frac{1}{2}\cos(x)\sin(y){\hat {\bf x}}-\frac{1}{2}\sin(x)\cos(y){\hat {\bf y}} \right], 
\end{eqnarray}
where
\begin{eqnarray}
G^+(z,t)=\int\limits_0^t f^+(\xi)f^-(\xi-2v_A\tau)\, d\tau.
\end{eqnarray}
In a similar fashion, we can derive the first-order correction ${\bf z}^-_1$, which is given by
\begin{eqnarray}
{\bf z}^-_1=G^-(z,t)\left[ -\frac{1}{2}\cos(x)\sin(y){\hat {\bf x}}+\frac{1}{2}\sin(x)\cos(y){\hat {\bf y}} \right], \label{first-}\\
G^-(z,t)=\int\limits_0^tf^+(\eta+2v_a\tau)f^-(\eta)\, d\tau, \label{gfirst-}
\end{eqnarray}
where $\eta=z-v_At$, $\tau=t$, and the integral in (\ref{gfirst-}) is done assuming $\eta=const$. We can proceed this way to the second and higher orders, calculating pressure corrections at each step. The expressions for the second-order corrections ${\bf z}_2^+$ and ${\bf z}_2^-$ are bulky; they will not be needed for our current discussion and will not be presented here. 

We can now study the evolution of residual energy after the interaction is switched on. Substituting the expansion (\ref{expansion}) into the expression for the residual energy, we find to the first nonvanishing order
\begin{eqnarray}
E_r=\langle {\bf z}^+\cdot {\bf z}^-\rangle = \langle {\bf z}_1^+\cdot {\bf z}_1^- \rangle= -\frac{1}{4}\langle G^+G^-\rangle,
\label{er}
\end{eqnarray} 
where the angular brackets mean averaging in the $x$, $y$, and $z$-directions. It can be checked that the following terms vanish: $\langle{\bf z}_0^+\cdot{\bf z}_0^- \rangle = \langle{\bf z}_1^\pm\cdot{\bf z}_0^\mp \rangle=\langle{\bf z}_2^\pm\cdot{\bf z}_0^\mp \rangle=0$. The averaging in formula (\ref{er}) can be easily performed since the functions $f^+$ and $f^-$ are not correlated with each other. We therefore derive:
\begin{eqnarray}
\langle G^+ G^-\rangle= \int \limits_0^t \langle f^+(\xi) f^+(\eta + 2v_A\tau)\rangle d\tau \int \limits_0^t \langle f^-(\eta) f^-(\xi - 2v_A\tau)\rangle d\tau .
\label{corr}
\end{eqnarray}
We now assume that the distributions of $f^+$ and $f^-$ are uniform along the $z$-direction, that is, their correlation functions depend only on the difference of their arguments: 
\begin{eqnarray}
\langle f^\pm(z_1) f^\pm(z_2)\rangle=\kappa^\pm(z_1-z_2). 
\label{kappa}
\end{eqnarray}
The correlation functions $\kappa^\pm$ are even, and they decline fast when their arguments exceed the correlation length $l_\|$. We also consider Fourier transforms of $f^\pm(z)$, $\kappa^\pm(z)$:
\begin{eqnarray}
f^\pm(z)=\frac{1}{2\pi}\int \limits_{-\infty}^{+\infty} {\tilde f}^\pm(k)\exp(ikz)dk, \quad \kappa^\pm(z)=\frac{1}{2\pi}\int \limits_{-\infty}^{+\infty} {\tilde \kappa}^\pm(k)\exp(ikz)dk,
\end{eqnarray}
so that their Fourier components are related as:
\begin{eqnarray}
\frac{1}{L}\langle |{\tilde f}^\pm(k)|^2 \rangle={\tilde \kappa}^\pm(k),
\label{fourier}
\end{eqnarray}
where $L$ is the system size in the $z$-direction ($L\gg l_\|$). Then the integrals in (\ref{corr}) can be easily done:
\begin{eqnarray}
\int \limits_0^t \langle f^+(\xi) f^+(\eta + 2v_A\tau)\rangle d\tau=\int \limits_0^t 
\kappa^+(\xi-\eta-2v_A\tau)d\tau =\int \limits_0^t \kappa^+(2v_At-2v_A\tau)d\tau ,\label{kappaint+} \\
\int \limits_0^t \langle f^-(\xi-2v_A\tau) f^-(\eta)\rangle d\tau=\int \limits_0^t 
\kappa^-(\xi-\eta-2v_A\tau)d\tau =\int \limits_0^t \kappa^-(2v_At-2v_A\tau)d\tau .
\label{kappaint-}
\end{eqnarray}
Now, we are interested in times that are much larger that the time necessary for each wave to shift by a correlation length, that is, we assume $t\gg l_\|/v_A$. In other words, we are interested in times exeeding a single collision time of two counter-propagating wave packets. Then, the integration limits  in (\ref{kappaint+}) and (\ref{kappaint-}) can be formally extended to infinity:
\begin{eqnarray}
\int \limits_0^t \kappa^\pm(2v_At-2v_A\tau)d\tau=\frac{1}{2v_A}\int \limits_0^\infty \kappa^\pm(z)dz = \frac{1}{4v_A}\int \limits_{-\infty}^{+\infty} \kappa^\pm(z)dz= \frac{1}{4v_A} {\tilde \kappa}^\pm(0).
\end{eqnarray}
According to (\ref{fourier}), this expression is positive. We therefore obtain that the residual energy is negative, and it is expressed through the wave correlation functions as:
\begin{eqnarray}
E_r=\langle {\bf z}_1^+\cdot {\bf z}_1^- \rangle=-\frac{1}{16v_A^2}{\tilde \kappa}^+(0){\tilde \kappa}^-(0)<0.
\label{er-final} 
\end{eqnarray}
This is the main result of our derivation. 
 
\section{Conclusions}
An important observation is that the residual energy is always negative, which means that the magnetic energy always exceeds the kinetic one. This result, obtained in the framework of incompressible MHD, provides a qualitative explanation for the solar wind observations. More general discussion indicates that this conclusion is quite robust, in that it is independent of the particular form of the initial $z^+$ and $z^-$ modes \cite{wang11}. In addition, direct numerical simulations indicate that it holds for strong turbulence as well.  \\

This work was supported   
by the US DoE Awards DE-FG02-07ER54932, DE-SC0003888, DE-SC0001794, the NSF Grant PHY-0903872, the NSF/DOE Grant AGS-1003451, 
and by the NSF Center for Magnetic Self-organization in Laboratory and Astrophysical Plasmas at the University of Wisconsin-Madison. High Performance Computing resources were
provided by the Texas Advanced Computing Center (TACC) at the
University of Texas at Austin under the NSF-Teragrid Project
TG-PHY080013N. SB is grateful to Mikhail Medvedev for useful discussions.  
\doingARLO{        
\begin{multimedia}
  \mmuid{523}
  \mmtype{gif}
  \mmsize{1.2Mb}
  \mmurl{http://yorktown.eng.yale.edu/test/msXXX/}
  \mmcaption{Fancy video}
  \label{fv}
\end{multimedia}}

\doingARLO[\bibliographystyle{aipproc}]
          {\ifthenelse{\equal{\AIPcitestyleselect}{num}}
             {\bibliographystyle{arlonum}}
             {\bibliographystyle{arlobib}}
          }


\begin{thebibliography}{22}
\expandafter\ifx\csname natexlab\endcsname\relax\def\natexlab#1{#1}\fi
\providecommand{\enquote}[1]{``#1''}
\expandafter\ifx\csname url\endcsname\relax
  \def\url#1{\texttt{#1}}\fi
\expandafter\ifx\csname urlprefix\endcsname\relax\def\urlprefix{URL }\fi
\providecommand{\eprint}[2][]{\url{#2}}

\bibitem[{Goldreich} and {Sridhar}(1995)]{goldreich_s95}
P.~{Goldreich}, and S.~{Sridhar}, \emph{Astrophys. J.} \textbf{438}, 763--775
  (1995).

\bibitem[{Biskamp}(2003)]{biskamp03}
D.~{Biskamp}, \emph{{Magnetohydrodynamic Turbulence}}, 2003.

\bibitem[{Podesta} et~al.(2007)]{podesta_rg07}
J.~J. {Podesta}, D.~A. {Roberts}, and M.~L. {Goldstein}, \emph{Astrophys. J.}
  \textbf{664}, 543--548 (2007).

\bibitem[{Tessein} et~al.(2009)]{tessein_etal09}
J.~A. {Tessein}, C.~W. {Smith}, B.~T. {MacBride}, W.~H. {Matthaeus}, M.~A.
  {Forman}, and J.~E. {Borovsky}, \emph{Astrophys. J.} \textbf{692}, 684--693
  (2009).

\bibitem[{Podesta} and {Bhattacharjee}(2009)]{podesta_b09}
J.~J. {Podesta}, and A.~{Bhattacharjee}, \emph{ArXiv e-prints}  (2009),
  \eprint{0903.5041}.

\bibitem[{Chen} et~al.(2011a)]{chen_etal10}
C.~H.~K. {Chen}, A.~{Mallet}, T.~A. {Yousef}, A.~A. {Schekochihin}, and T.~S.
  {Horbury}, \emph{MNRAS} \textbf{415}, 3219 (2011a).

\bibitem[{Borovsky}(2011)]{borovsky11}
J.~E. {Borovsky}, \emph{submitted to J Geophys. Res.}  (2011).

\bibitem[{Chen} et~al.(2011b)]{chen_etal11}
C.~H.~K. {Chen}, S.~D. {Bale}, C.~{Salem}, and F.~S. {Mozer}, \emph{Astrophys. J} \textbf{737}, L41 (2011b).

\bibitem[{Roberts}(2010)]{roberts10}
D.~A. {Roberts}, \emph{Journal of Geophysical Research (Space Physics)}
  \textbf{115}, A12101 (2010).

\bibitem[{Borovsky}(2010)]{borovsky10}
J.~E. {Borovsky}, \emph{Physical Review Letters} \textbf{105}, 111102 
  (2010).

\bibitem[{Li} et~al.(2011)]{li_etal11}
G.~{Li}, B.~{Miao}, Q.~{Hu}, and G.~{Qin}, \emph{Physical Review Letters}
  \textbf{106}, 125001 (2011).

\bibitem[{Boldyrev} et~al.(2011)]{boldyrev_pbp11}
S.~{Boldyrev}, J.~C. {Perez}, J.~E. {Borovsky}, and J.~J. {Podesta},
  \emph{ArXiv e-prints}  (2011), \eprint{1106.0700}.

\bibitem[{Boldyrev} and {Perez}(2009)]{boldyrev_p09}
S.~{Boldyrev}, and J.~C. {Perez}, \emph{Phys. Rev. Lett.} \textbf{103},
  225001 (2009).

\bibitem[{Wang} et~al.(2011)]{wang11}
Y.~{Wang}, S.~{Boldyrev}, and J.~C. {Perez}, \emph{ArXiv e-prints}  (2011),
  \eprint{1106.2238}.

\bibitem[{Grappin} et~al.(1983)]{grappin_pl83}
R.~{Grappin}, A.~{Pouquet}, and J.~{Leorat}, \emph{Astron. Astrophys.}
  \textbf{126}, 51--58 (1983).

\bibitem[{Zank} et~al.(1996)]{zank_ms96}
G.~P. {Zank}, W.~H. {Matthaeus}, and C.~W. {Smith}, \emph{J. Geophys. Res.}
  \textbf{101}, 17093--17108 (1996).

\bibitem[{M{\"u}ller} and {Grappin}(2005)]{muller_g05}
W.~{M{\"u}ller}, and R.~{Grappin}, \emph{Phys. Rev. Lett.} \textbf{95},
  114502 (2005).

\bibitem[{Marsch} and {Mangeney}(1987)]{marsch_m87}
E.~{Marsch}, and A.~{Mangeney}, \emph{J. Geophys. Res.} \textbf{92}, 7363--7367
  (1987).

\bibitem[{Ng} and {Bhattacharjee}(1996)]{ng_b96}
C.~S. {Ng}, and A.~{Bhattacharjee}, \emph{Astrophys. J.} \textbf{465}, 845
  (1996).

\bibitem[{Galtier} et~al.(2000)]{galtier_nnp00}
S.~{Galtier}, S.~V. {Nazarenko}, A.~C. {Newell}, and A.~{Pouquet},
  \emph{Journal of Plasma Physics} \textbf{63}, 447--488 (2000).

\bibitem[{Galtier} and {Chandran}(2006)]{galtier_c06}
S.~{Galtier}, and B.~D.~G. {Chandran}, \emph{Phys. Plasmas} \textbf{13},
  114505 (2006).

\bibitem[{Perez} and {Boldyrev}(2009)]{perez_b09}
J.~C. {Perez}, and S.~{Boldyrev}, \emph{Phys. Rev. Lett.} \textbf{102},
  025003 (2009).

\end{thebibliography}

\end{document}